\documentclass[doublecol]{epl2} 
\usepackage{url}
\usepackage{graphicx}
\usepackage{float}
\usepackage{color}
\usepackage{amssymb}
\usepackage{amsmath} 
\usepackage[T1]{fontenc}
\usepackage{comment} 
\usepackage{geometry} 
\usepackage{bm}
\usepackage{cite}
\usepackage[colorlinks,bookmarks=false,citecolor=blue,linkcolor=blue,urlcolor=blue]{hyperref} 
\title{Dirac cones and higher-order topology in quasi-continuous media}

\author{Zhi-Kang Lin\inst{1} \and Jian-Hua Jiang\inst{1}\thanks{E-mail: \email{jianhuajiang@suda.edu.cn} }}
\shortauthor{Zhi-Kang Lin \etal}

\institute{                    
  \inst{1} School of Physical Science and Technology \& Collaborative Innovation Center of Suzhou Nano Science and Technology, Soochow University, Suzhou 215006, China
}


\abstract{We consider the Dirac cones and higher order topological phases in quasi-continuous media of classical waves (e.g., photonic  and sonic crystals). Using sonic crystals as prototype examples, we revisit some of the known systems in the study of topological acoustics. We show the emergence of various Dirac cones and higher order topological band gaps in the same motherboard by tuning the geometry of the system. We provide a pedagogical review of the underlying physics and methodology via the bulk-edge-corner correspondence, symmetry-based indicators, Wannier representations, filling anomaly, and fractional corner charges. In particular, the theory of the Dirac cones and the higher-order topology are put in the same framework. These examples and the underlying physics principles can be inspiring and useful in future study of higher order topological metamaterials.
}
\begin{document}

\maketitle

\section{Introduction}
The past few years have witnessed the rapid development in a new frontier of topological phases of matter \cite{hasan2010colloquium,qi2011topological} that is termed as the higher-order topological insulators (HOTIs) \cite{schindler2018higher}. HOTIs are firstly indicated by the existence of chiral edge states in three-dimensional (3D) axion insulators in magnetic fields \cite{sitte2012topological,zhang2013surface} and then intrigue tremendous interest due to the celebrated multipole insulators \cite{benalcazar2017quantized,benalcazar2017electric}. At the same time or later, several concrete models and theories have being proposed \cite{langbehn2017reflection,song2017d,ezawa2018higher,ezawa2018minimal,ezawa2018topological,geier2018second,schindler2018higher,khalaf2018symmetry,van2018higher,khalaf2018higher,hsu2018majorana,matsugatani2018connecting,franca2018anomalous,benalcazar2019quantization,liu2019shift,fang2019new,trifunovic2019higher,ahn2019stiefel,okuma2019topological,kooi2019classification,park2019higher,kudo2019higher,queiroz2019splitting,yang2020type,araki2020z,roberts2020second,rasmussen2020classification,peterson2020fractional,khalaf2021boundary,trifunovic2021higher,xie2021higher}. Prototypes of HOTIs include the two-dimensional (2D) second-order topological insulators which host one-dimensional (1D) gapped edge states and zero-dimensional (0D) in-gap topological corner states. Here, the term ``second-order'' means the protected topological boundary states have a co-dimension of two (i.e., the topological boundary states have two dimensions lower than the bulk states). Such HOTIs generalize the conventional bulk-edge correspondence to the bulk-edge-corner correspondence and unveil a unprecedented regime of multidimensional topological physics with versatile topological phenomena. Unlike the conventional topological insulators which are protected by the time-reversal symmetry, HOTIs are protected by the crystalline symmetry and characterized by the bulk topological invariants such as the quantized multipole polarizations \cite{benalcazar2017quantized,benalcazar2017electric}. HOTIs represent a large category of topological crystalline insulator phases which widely emerge in  natural~\cite{schindler2018higher,2018Higher,2019Higher,yue2019symmetry,xu2019higher,ezawa2019second,choi2020evidence,lee2020two,aggarwal2021evidence,noguchi2021evidence} and artificial~\cite{peterson2018quantized,serra2018observation,imhof2018topolectrical,xue2019acoustic,ni2019observation,bao2019topoelectrical,zhang2019second,xie2018second,noh2018topological,ota2019photonic,xie2019visualization,chen2019direct,fan2019elastic,kempkes2019robust,zhang2019dimensional,mittal2019photonic,xue2019realization,el2019corner,pelegri2019second,zhang2019deep,liu2020octupole,li2020higher,zhang2020higher,zhou2020twisted,meng2020realization,yang2020gapped,zhang2020symmetry,qi2020acoustic,weiner2020demonstration,zheng2020three,he2020quadrupole,ni2020demonstration,xue2020observation,chen2020plasmon,wang2020higher,yang2020observation,wakao2020higher,banerjee2020coupling,xiong2020corner,wu2021chip,zhang2021valley,chen2021corner} materials. In many cases, the nonzero polarizations in HOTIs manifest the filling anomaly~\cite{benalcazar2019quantization} of the bulk states in finite systems, which, from a real-space perspective, can be revealed via the Wannier centers of the Bloch bands \cite{benalcazar2019quantization,khalaf2021boundary}. 

What makes HOTIs particularly attractive for material scientists are their rich topological phenomena across dimensions.  Topological phenomena at different dimensions in the same material could provide multiplexing applications. For instance, the gapped surface states in three-dimensional HOTIs can provide surface states with high density of states to accelerate surface chemical reaction processes. Meanwhile, the bulk, through filling anomaly, gives rise to high-density surface charges. In addition, the gapless hinge states offer robust transport on 1D hinge channels. In photonic HOTIs, it has been shown that edge states can provide 1D waveguide channels \cite{kim2020recent}, while the 0D corner states can serve as robust cavities that may enable ultra-low threshold lasing \cite{zhong2021theory,zhang2020low,kim2020multipolar,han2020lasing}.

\begin{figure*}[t]
\centering\includegraphics[width=0.8\linewidth]{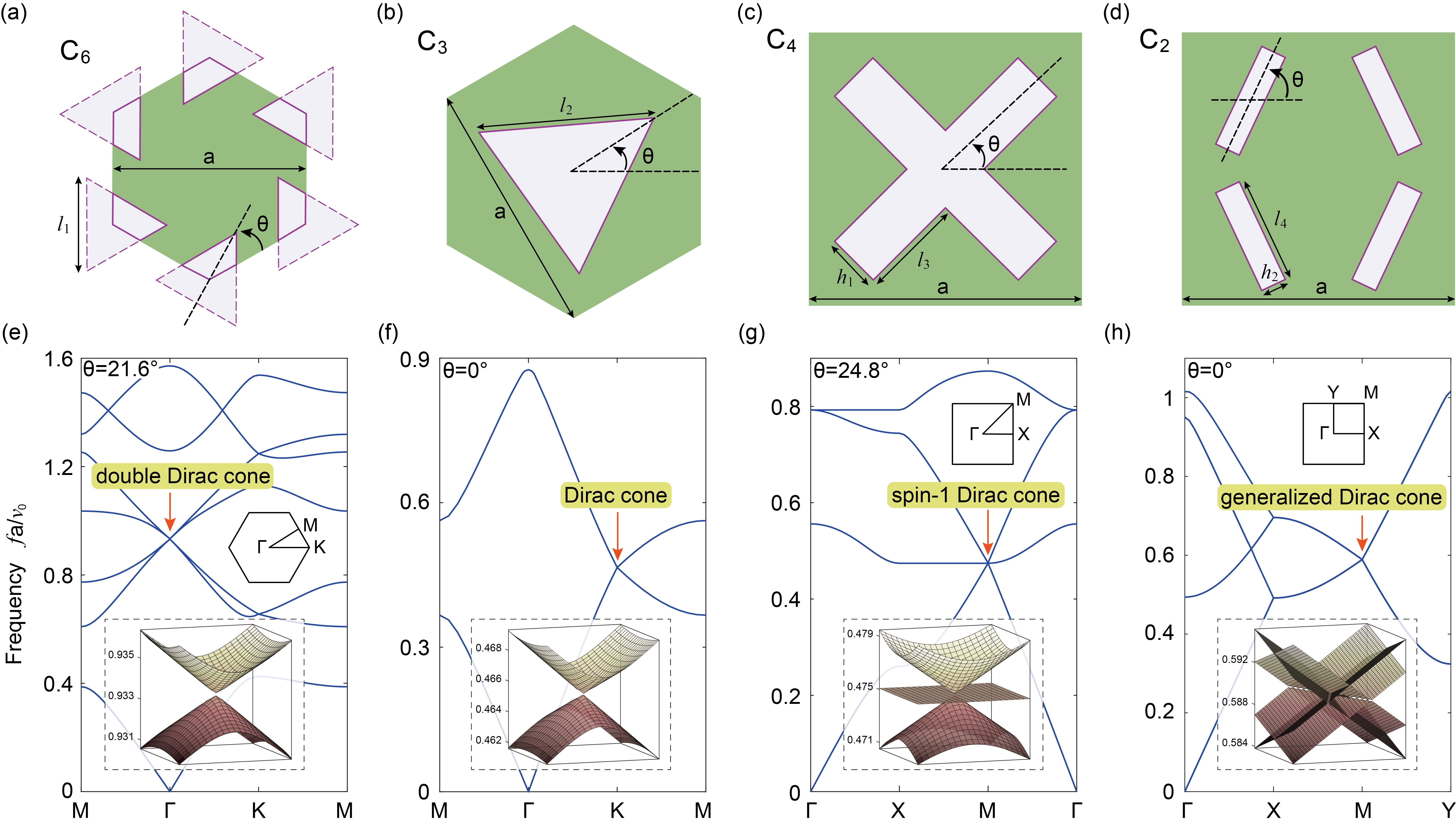}
\caption{ Acoustic analogue of various Dirac cones in SCs with (a) sixfold ($C_6$), (b) threefold ($C_3$), (c) fourfold ($C_4$) and (d) twofold ($C_2$) rotation symmetries. Only the unit cells of the SCs are illustrated where the  rotatable acoustic scatterers and the air regions are denoted by the white and green regions, respectively. Except for the rotation angle $\theta$ of scatterers, the other geometric parameters  $l_1=0.49a$ in (a), $l_2=0.89a$ in (b), $l_3=0.375a$, $h_1=0.2a$ in (c) and $l_4=0.4a$, $h_2=0.1a$ in (d) are fixed. Here, $a$ denotes the lattice constant for all cases. (e)-(f): The corresponding acoustic band structures for SCs in (a)-(d) with a double Dirac cone at the $\Gamma$ point (e), a Dirac cone at the $K$ point (f), a spin-1 Dirac cone at the $M$ point (g) and a generalized Dirac cone at the $M$ point (h). The respective rotation angle $\theta$ is listed  on the top-left corner of each Figure. Insets show the Brillouin zone and the 3D band dispersion around the degenerate points.}
\label{Fig.1}
\end{figure*}

On the other hand, Dirac cones \cite{neto2009electronic} that emerge in various photonic and phononic artificial materials (such as photonic crystals and phononic crystals) have been the focus of researches in the study of metamaterials, because of their unique properties.
For instance, in photonic crystals, Dirac cones can be utilized to realize all-dielectric zero refractive index medium that has low dissipation and enables extraordinary manipulation of light \cite{huang2011dirac,sakoda2012double,chan2012dirac,li2021dirac}. 

Since there are plenty of studies on how to construct HOTIs from tight-binding models, here we emphasize the construction of HOTIs in the Bragg scattering regime. In this regime, which we denote as the quasi-continuous regime, waves are nearly free, while the Bloch bands are formed by Bragg scatterings. Examples include photonic crystals, sonic crystals (SCs) and phononic crystals (including phonons in natural and artificial solid lattice systems), as well as some other metamaterials. In such a regime, although there is no tight-binding picture, HOTIs can still be realized and manipulated by tuning the geometry of the scatterers in each unit-cell. Here, by revisiting some of the recently studied acoustic systems, we show that Dirac cones and HOTIs can be realized in the same system in quasi-continuous media. We further illustrate how to obtain the topological invariants (i.e., the symmetry indicators of the Bloch bands, the filling anomaly from the Wannier orbital picture, and the fractional corner charges) that characterize the higher-order topology of the Bloch bands.

\section{Dirac cones in 2D SCs}

We specifically consider four 2D airborne SCs with  $C_n$ rotation symmetries ($n = 2, 3, 4, 6$) as illustrated in Figs.~\ref{Fig.1}(a)-(d). The rotation symmetries play an essential role in the theory of higher-order topological phases \cite{benalcazar2019quantization,van2018higher,song2017d,benalcazar2017electric,fang2019new}. In these SCs, the band inversion at one of the high symmetry points (HSPs) can be triggered by tuning the rotation angles of the scatterers $\theta$. Such band inversions lead to topological transitions and the emergence of various Dirac cones.

Here, we focus on the 2D airborne SCs where the acoustic scatterers are placed in the air background. The wave equation for the acoustic wave is given in details in Supplementary information. The scatterers here are realized by photosensitive resins via the commercial 3D printing technology. As depicted in Figs.~\ref{Fig.1}(a)-(d), four designed SCs are introduced, which separately enjoy the sixfold ($C_6$), fourfold ($C_4$), threefold ($C_3$) and twofold ($C_2$) rotation symmetries. 

It is found that at certain angles, those SCs host various Dirac cones [see Figs.~\ref{Fig.1}(e)-(h)]. For instance, a double Dirac cone with fourfold degeneracy can appear in the $C_6$-symmetric SC at the Brillouin zone center. At each corner of the hexagonal unit-cell, there is a scatterer made of epoxy [see Fig.~\ref{Fig.1}(a)], which is an equilateral triangle with the side length $l_1=0.49a$. While keeping the $C_6$ rotation symmetry and rotating the scatterers, the acoustic bands can be tuned. At the rotation angle $\theta=21.6^\circ$, a double Dirac cone emerges at the Brillouin zone center [Fig.~\ref{Fig.1}(e)]. Such a double Dirac cone is initially studied for zero-index metamaterials \cite{sakoda2012double} and then plays an pivotal role in the discovery of the analogous spin Hall effect in various photonic, and phononic systems \cite{wu2015scheme,he2016acoustic}. 

The SC with the $C_3$ rotation symmetry in Fig.~\ref{Fig.1}(b) is known for the acoustic realization of the conventional~\cite{lu2016valley,lu2017observation} and higher-order ~\cite{zhang2021valley} valley Hall insulators. In these SCs, there is a single epoxy scatterer at the center of the unit-cell. This scatterer is an equilateral triangle with the side length $l_2=0.89a$. At $\theta=0^\circ$, Dirac cones emerge at the $K$ and $K^\prime$ points [see Fig.~\ref{Fig.1}(f)]. 

The $C_4$-symmetric SC in Fig.~\ref{Fig.1}(c) simulates an acoustic Lieb lattice. In a unit-cell, there is a single epoxy scatterer at the center. This scatterer is designed to be a cross shape characterized by the length $l_3=0.375a$ and the width $h_1=0.2a$. By tuning the rotation angle to $\theta=24.8^\circ$, a spin-1 Dirac cone with threefold degeneracy, emerges at the $M$ point [Fig.~\ref{Fig.1}(g)]. The spin-1 Dirac cone is composed of two linear bands intersecting with a nearly flat band and has been studied with great interest as a candidate for double-zero index metamaterials \cite{huang2011dirac,chan2012dirac,dubois2017observation,xu2020three,li2021dirac}.

The $C_2$-symmetric SC in Fig.~\ref{Fig.1}(d) is originally proposed and experimentally realized as a second-order topological insulator \cite{zhang2019second}. 
There are four scatterers in each unit-cell. The topological transition takes place at $\theta=0^\circ$ where a generalized Dirac cone with fourfold degeneracy emerges at the $M$ point. We use the term of ``generalized Dirac cone'' here due to the existence of another nodal lines crossing it. We remark that the linear dispersion of all the Dirac cones discussed above can be illustrated by the $k\cdot p$ method near the degenerate points\cite{mei2012first}.

\begin{figure*}[t]\centering\includegraphics[width=0.9\linewidth]{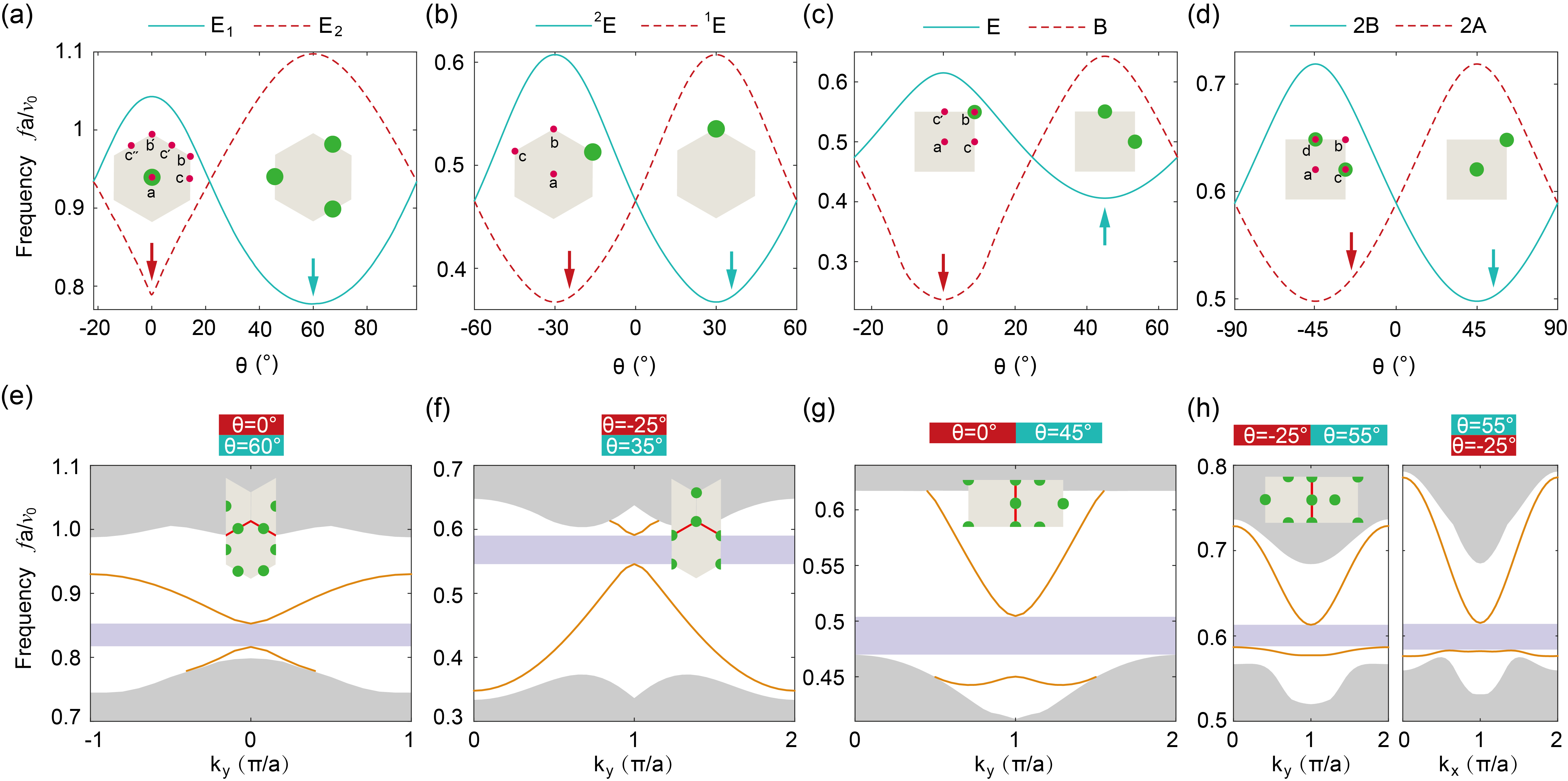}
\caption{Schematic of the topological phase transitions of bulk states and the emergent gapped edge states. (a)-(d) The topological phase diagrams of bulk states for the SCs in Figs.~\ref{Fig.1}(a)-(d). The frequencies of the concerned states with different symmetry representations versus the rotation angle are plotted as the red dashed and green solid curves, respectively. Separately, the symmetry representations are  $E_1$ and $E_2$ in (a), $^2E$ and $^1E$ in (b), $E$ and $B$ in (c) and $B$ and $A$ in (d), as labelled on the top of each Figure. The labels of the representations are based on the character tables on the Bilbao Crystallographic Server \cite{timmurphy.org}. Insets illustrate the location of real-space Wannier centers of the lower-frequency Bloch bands below the opened band gap. The red dots denote the Wyckoff positions. (e)-(h) The corresponding projected band structures along the wavevectors calculated from the ribbon-shaped structures as sketched on the top of each Figure. The domain walls in one direction are between two topologically distinct phases with specific rotation angles. The periodic boundary condition remains in the other direction. The orange curves denote the gapped edge states whose wave functions localize at the domain walls. The blue regions represent the frequency ranges of the edge band gaps. Insets schematically illustrate the Wannier centers around the domain-wall boundaries.}
\label{Fig.2}
\end{figure*}

\section{Topological transitions and edge states}
We note that the periodicity of the rotation angle is $120^\circ$ for both the $C_6$- and $C_3$-symmetric SCs, while $90^\circ$ and $180^\circ$ for the $C_4$- and $C_2$-symmetric SCs, respectively. By tuning the angle $\theta$ while preserving the rotation symmetries, the band gaps can be closed and reopen, leading to the topological transitions. This process is well captured by the flip of the eigen-frequencies of the Bloch states with different symmetry representations, as shown in Figs.~\ref{Fig.2}(a)-(d).  

Specifically, for the $C_6$-symmetric SC, the parity inversion occurs between two doubly degenerate representations $E_1$ and $E_2$ of the $C_6$ point group. The parity inversion leads to the emergence of the double Dirac cone at the $\Gamma$ point at $\theta=21.6^\circ$ for the parameters adopted here.

For the $C_3$-symmetric SC, tuning the rotation angle $\theta$ can trigger the inversion between the nondegenerate representations $^1E$ and $^2E$ at the $K$ and $K^\prime$ points. The transition leads to the emergence of the Dirac cones at $\theta=0^\circ$ where the $^1E$ and $^2E$ representations become degenerate. 

In the case of the $C_4$-symmetric SC, by tuning the angle $\theta$, inversion of the doubly degenerate representation $E$ and the nondegenerate representation $B$ can be triggered at the Brillouin zone corner (i.e., the $M$ point). At the transition point, $\theta=24.8^\circ$, the $E$ and $B$ representations are degenerate, leading to a triply-degenerate Dirac cone.

The $C_2$-symmetric SC experiences a parity inversion at the $M$ point when the rotation angle $\theta$ is tuned. This is associated with the flip of the doubly degenerate $2B$ and $2A$ representations. At the transition point, the degeneracy of the $2B$ and $2A$ representations leads to the generalized Dirac cone at the $M$ point.

The band topology of the $C_n$-symmetric insulators can be distinguished by the rotation symmetry representations of the bulk Bloch states at the HSPs in the Brillouin zone, which contribute to several topological indices \cite{benalcazar2019quantization}. The topological indices of different phases for the four SCs are provided in details in Supplementary Information.


We present in Figs.~\ref{Fig.2}(e)-(f) the calculated acoustic edge states. The corresponding ribbon-shaped supercells are schematically depicted on the top of each figure, where the domain wall is between two topologically distinct SCs with specific rotation angles. Specifically, the zigzag domain-wall boundaries are formed in the $C_6$- and $C_3$-symmetric cases. The projected bands for all cases exhibit  two edge states in the band gap, which are gapped because there is no symmetry at the edge that protects them to be gapless.  Finally, we remark that the edge band gap can be closed for the $C_3$- and $C_2$-symmetric SCs, when the edge boundary has the emergent glide symmetry, as shown in Refs.~\cite{zhang2019second,xiong2020corner}. The edge gap closing leads to an independent topological transition at the edges, which is not related with the bulk topological transitions.

\begin{figure*}[t]
\centering\includegraphics[width=0.9\linewidth]{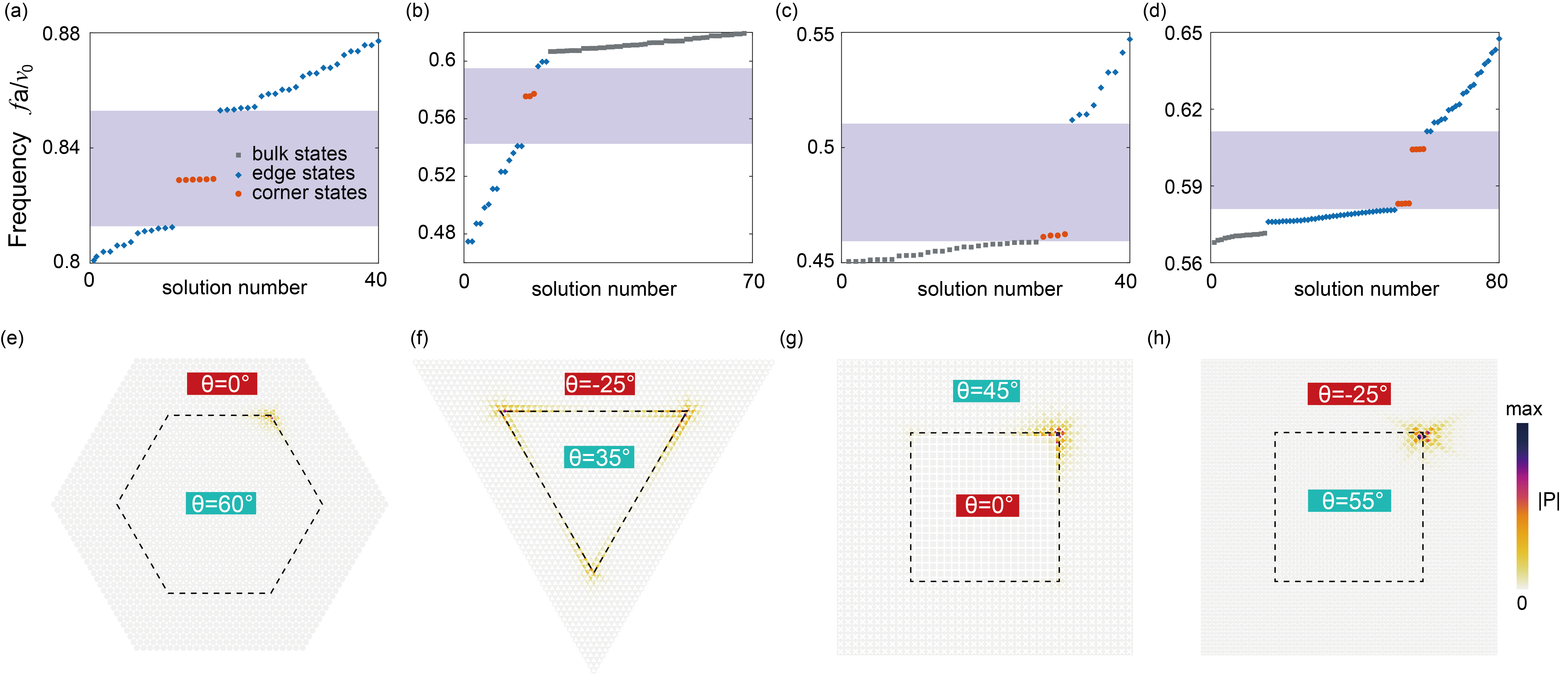}
\caption{Corner states as a direct demonstration of the higher-order band topology. (a)-(d) The corresponding eigen spectra of the finite box-shaped structures in (e)-(h). The corner states emerge in the edge band gaps. The bulk, edge and corner states are denoted by the grey, blue and red dots, respectively. (e)-(f) The profiles of the acoustic pressure fields of one corner state for the $C_6$, $C_3$, $C_4$ and $C_2$-symmetric cases, respectively.  The profiles show the localization at corners. The rotation angles of inner and outer SCs are labelled in Figures.}
\label{Fig.3}
\end{figure*}

\section{Wannier representations} 
In the Wannier-representable HOTIs studied in this work, the emergence of the edge states can be associated with the real-space Wannier orbitals. In other words, the two branches of the edge states can be understood as evolved from the two Wannier orbitals exposed at the edge boundary. The band representation theories give the $C_n$-symmetric Wannier orbitals located at the HSPs of the unit cell (i.e., Wyckoff positions) according to the ``band representations'' (i.e., the symmetry representations of the Bloch bands). The band representations  are listed in details in the Bilbao Crystallographic Server \cite{timmurphy.org1}. The Wannier centers for different gapped phases of the four SCs are illustrated in the insets of Figs.~\ref{Fig.2}(e)-(f).

Specifically, for the $C_6$-symmetric SC, there are always three Bloch bands below the gap. The three corresponding Wannier centers are at the center of the unit cell (i.e., the Wyckoff position $a$) if $\theta\in(-21.6^\circ,21.6^\circ)$, or the edges of the unit-cell (i.e., the Wyckoff positions $c$,$c'$ and $c''$) if $\theta\in(21.6^\circ,98.4^\circ)$. The former Wannier configuration corresponds to the trivial atomic insulator, while the latter corresponds to an obstructed atomic insulator. For the $C_3$-symmetric case, only one Bloch band is below the concerned band gap. The single Wannier center is at the Wyckoff position $c$ if $\theta\in(-60^\circ,0^\circ)$, or at the Wyckoff position $b$ if $\theta\in(0^\circ,60^\circ)$. Such two gapped phases are both topological, since for both of them the Wannier center is away from the unit-cell center. For the $C_4$-symmetric SC, there is one band below the gap if $\theta\in(-24.8^\circ,24.8^\circ)$. However, if $\theta\in(24.8^\circ,65.2^\circ)$, there are two Bloch bands below the gap. In the former case, the Wannier center is at the corner of the unit cell (i.e., the Wyckoff position $b$), whereas in the latter case, the Wannier centers are at the edge centers of the unit-cell (i.e., the Wyckoff positions $c$ and $c'$). Both gapped phases are topological, but of distinct properties. For the $C_2$-symmetric SC, there are always two bulk bands below the gap. If $\theta\in(-90^\circ,0^\circ)$, the two Wannier centers are at the edge centers. In comparison, if $\theta\in(0^\circ,90^\circ)$, one of the Wannier center is at the unit-cell center, while the other is at the unit-cell corner. 

For all the cases, there are two Wannier centers exposed to the edge boundary within an edge supercell when two SCs with topologically distinct band gaps are placed together. This is schematically illustrated in the insets in Figs.~\ref{Fig.2}(e)-(f) where the domain-wall boundaries are marked with red lines. The two Wannier orbitals exposed to the edge boundary are responsible for the emergence of the edge states~\cite{benalcazar2019quantization}.

\begin{figure*}[t]
\centering\includegraphics[width=0.8\linewidth]{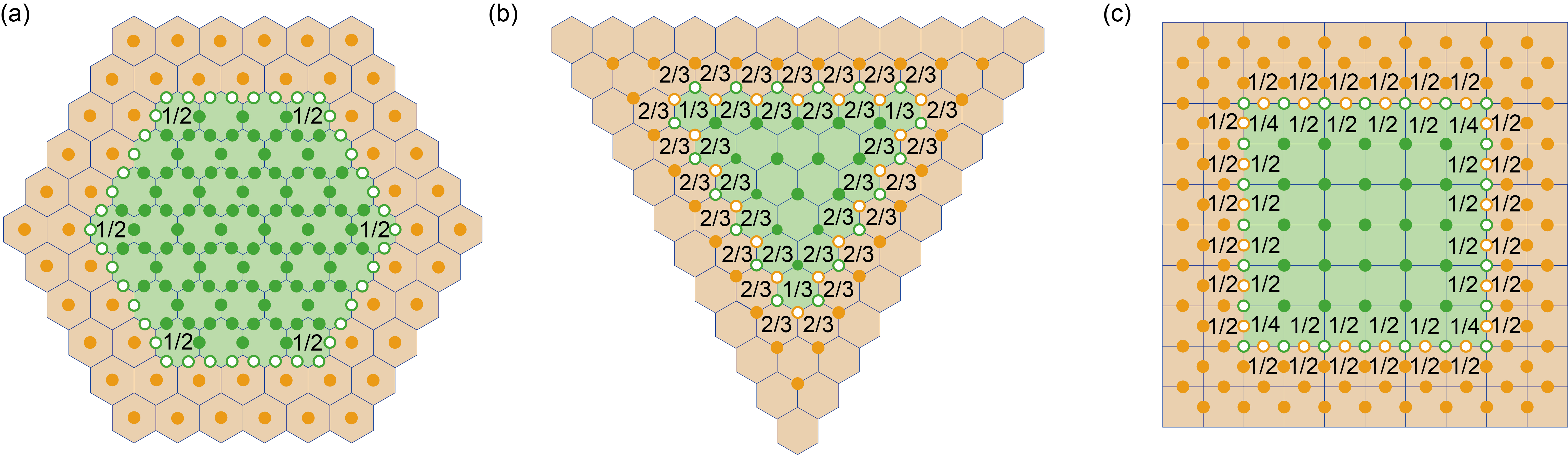}
\caption{Fractional edge and corner charges. The Wannier centers of inner and outer SCs in finite box-shaped structures are denoted by green and orange dots, respectively. The hollow dots represent the Wannier centers occupied at the domain wall between inner and outer SCs. In (a), for the $C_6$-symmetric case, one unit cell at each corner manifests the $\frac{1}{2}$ charge. Similarly, the fractional edge and corner charges are labelled for unit cells in (b) for the $C_3$-symmetric case and in (c) for  the $C_2$-symmetric case. The $C_4$-symmetric case is similar to the $C_2$-symmetric case expect for one more Wannier centers occupied at the unit-cell center of inner SCs which denote no fractional charges.}
\label{Fig.4}
\end{figure*}

\section{Higher-order topology, corner states and fractional corner charges}

The corner states may appear in the edge band gap as a direct manifestation of the higher-order band topology. For all the four cases, the edge and corner boundaries are formed in the finite-sized systems, as shown in Figs.~\ref{Fig.3}(e)-(h). The rotation angles of the inner and outer SCs are chosen as the same as those in Fig.~\ref{Fig.2}. Specifically, the topological SCs are placed in the inner region for the $C_6$-symmetric case. For the other three cases, the Wannier centers of the inner SCs, respectively, reside at the Wyckoff position $b$ for the $C_3$-symmetric case and the unit-cell corner for both the $C_4$- and $C_2$-symmetric cases. 

The calculated eigen-spectra for these finite systems are presented in Figs.~\ref{Fig.3}(a)-(d) which manifest the bulk-edge-corner correspondence with the corner states emerging in the edge band gap. The localization of the corner states are confirmed by the acoustic pressure profiles in Figs.~\ref{Fig.3}(e)-(h). We remark that these features remain intact if the inner SC exchanges with the outer SC for each configuration. Note that in those calculations, the hard-wall boundary condition is set at the outer most boundary, to keep the full system closed and Hermitian.

In addition to the above spectral signatures, which although in many cases are not fragile but strictly speaking not completely protected by the higher-order band topology, there are two robust bulk properties protected by the band topology: filling anomaly and fractional corner charges. These two properties are interrelated with each other. In simple words, filling anomaly refers to the phenomena that in a finite system, the number of bulk eigenstates is different from the number of Wannier orbitals in the system. The filling anomaly is primarily due to the Wannier orbitals occupied at the edge or corners. A direct consequence of the filling anomaly is the emergence of the fractional charges at the edge and corner boundaries. 


Following Ref.~\cite{benalcazar2019quantization}, the corner charges are formulated based on the topological indices, as given in Supplementary information. However, they rely on the specific shapes of the finite-sized systems \cite{benalcazar2019quantization}. Instead, it is  convenient to utilize the Wannier orbital distribution to detect the fractional edge and corner charges. As depicted in Fig.~\ref{Fig.4}, the Wannier centers of the inner and  outer regions are denoted by the green and orange dots, respectively.  Specially, the Wannier centers occupied at the edges and corners are denoted by the hollow dots. We consider the charges of each unit cell, which can be inferred by counting the number of Wannier centers dropped in the unit cell. For instance, as shown in  Fig.~\ref{Fig.4} (a) for the $C_6$-symmetric SC, the single unit cell in the inner bulk region has six half of the Wannier centers and hence carries the $\frac{1}{2}\times6=3$ charges. At the corner unit cell, the hollow dots denote no charges and therefore the whole charges are  $\frac{1}{2}\times3\ mod\ 1=\frac{1}{2}$. The same derivation can apply to the $C_3$-, $C_4$- and $C_2$-symmetric cases, their distributions of the fractional charges are presented in  Figs.~\ref{Fig.4} (b) and (c). The fractional charges around the corners are the primary cause of the corner states. Note that there is no edge fractional charge  in the $C_6$-symmetric case, while the edge states still emerge in the bulk band gap. Such a feature indicates that the filling anomaly may not manifest as the fractional charges. In contrary, the fractional charges at the edge or corner always indicate the filling anomaly. It's also worth noting that the the fractional charges calculated from the Wannier centers are theoretical values in the limit with extremely localized Wannier orbitals. The real charges are approximate values compared to them, which can be calculated from the first-principle numerical method, as elaborated in Supplementary information.


\section{Conclusions and outlook}
In this paper, we revisit the higher-order topological phases in quasi-continuous media. Using SCs as prototype examples, we show that the Dirac cones and the higher-order topological insulator phases can emerge in the same motherboard of SCs, upon tuning the geometry of the SCs. In particular, we review the underlying physics and theoretical methods used in the literature to analyze the higher-order topology and Dirac cones in the a unified framework, i.e., the symmetry-based analysis of the Bloch bands and the underlying Wannier representations. In addition to the known spectral bulk-edge-corner correspondence, we emphasize that the higher-order topological insulator phases gives rise to filling anomaly and fractional charges at the edge and corner boundaries. The results presented in this paper is useful for future studies on quasi-continuous topological metamaterials. 

We remark that in most cases the phononic bands do not have chiral symmetry. In some extreme cases, the edge and corner states can be destroyed by the absence of the chiral symmetry. However, in most cases, the corner states and the edge states remain robust in the bulk band gap. On the other hand, although it is not easy to probe the fractional corner changes, using photonic analogs people are now able to measure the fractional corner charges. New ideas of probing the higher-order topology and filling anomaly were also proposed and realized very recently (e.g., topological Wannier cycles \cite{lin2021experimental,lin2021topological}.)

Finally, we would like to mention that there are also fragile topological insulators which are not Wannier representable but also supports the spectral bulk-edge-corner correspondence, filling anomaly, fractional edge and corner charges. From the experimental side, it is hard to distinguish the difference between HOTIs and fragile topological insulators. Finally, it would be very interesting as well to study the non-Hermitian regime for the higher-order and Dirac cone phases. Due to their tunability and subwavelength nature as well as compatibilities for various applications, quasi-continuous media are expected to play an important role in future study and applications of higher-order topological metamaterials.

 \acknowledgments

\end{document}